# 100ps time resolution with thin silicon pixel detectors and a SiGe HBT amplifier.

Mathieu Benoit[a], Roberto Cardarelli[b], Stéphane Débieux[a], Yannick Favre[a], Giuseppe Iacobucci[a], Marzio Nessi[a,c], Lorenzo Paolozzi[a,c]*and Kenji Shu[d].

[a] *DPNC, University of Geneva,*
  *CH 1211 Geneva 4, Switzerland*

[b] *INFN, Sezione di Roma Tor Vergata,*
  *Via della ricerca scientifica 1 Roma, Italy*

[c] *CERN*
  *Geneva, Switzerland*

[d] *CERN Summer student from The University of Tokyo,*
  *7-3-1 Hongo, Bunkyo-ku, Tokyo 113-0033, Japan.*
  *E-mail:* `lorenzo.paolozzi@cern.ch`

ABSTRACT: A 100μm thick silicon detector with 1mm$^2$ pad readout optimized for sub-nanosecond time resolution has been developed and tested. Coupled to a purposely developed amplifier based on SiGe HBT technology, this detector was characterized at the H8 beam line at the CERN SPS. An excellent time resolution of (106±1)ps for silicon detectors was measured with minimum ionizing particles.

---

* Corresponding author.

# Contents

# 1. Introduction.

## 1.1 Design of a sub-nanosecond time resolution front end electronics for silicon sensors

Silicon detectors are commonly used to realize tracking systems with micrometric space resolution or for spectroscopic measurements, disregarding their potential excellent time resolution. In order to achieve sub-nanosecond time resolution, intrinsic fluctuations of the charge collection time inside the sensor must be minimized. This can be achieved by designing a pixel geometry with a uniform weighting field and by saturating the carrier drift velocity in the whole sensor volume. Under these assumptions the time response jitter of the detector, which is the time fluctuation of the production of the charges in the sensor, is well below the picosecond scale.

Two intrinsic effects that lead to a degradation of the detector time resolution should however be taken into account: the time fluctuations due to the noise of the output signal and the variation of the geometrical distribution of the deposited energy inside the sensor volume.

The first effect depends mostly on the preamplifier noise (N), whose contribution to the time resolution $\sigma_t$ is given by:

$$\sigma_t = \frac{\sigma_V}{\left(\frac{dV}{dt}\right)} \cong \frac{t_{rise}}{(S/N)},$$

where $\sigma_V$ is the voltage RMS noise of the amplifier and $\frac{dV}{dt}$, $t_{rise}$ and $S/N$ are respectively the output pulse slope, rise time and signal to noise ratio.

In the case of a charge-integrating amplifier, the rise time of the output pulse $t_{rise}$ is equal to the charge collection time, and the above formula can be written as:

$$\sigma_t \cong \frac{N}{v_s \frac{dQ}{dx}}$$

where $v_s$ is the saturated drift velocity of the carriers and $\frac{dQ}{dx}$ is the charge produced per unit length. From this expression it can be shown that for a single layer detector [1] it is not possible to improve the time resolution by changing its thickness and the only possible improvement can be obtained by the reduction of the amplifier noise.

The second effect that could lead to a degradation of the intrinsic time resolution of the detector is the charge-collection noise [2]. This is an equivalent noise on the induced current, which is generated by the granularity and size difference of primary charge clusters produced by the interacting particle. As long as all the clusters move between the electrodes, a constant current is induced in the read-out line; as the first cluster reaches the electrode, its charge contribution is removed from the signal, thus producing an irreducible equivalent noise charge (ENC) only present during the charge collection, whose contribution to time resolution can be minimized to less than 20ps by reducing the detector thickness [2].

Starting from the above considerations, we have designed an amplifier with very low noise for a fast pulse shaping and capable of operating with an input capacitance of the order of 1pF. Our choice was to use silicon-germanium Heterojunction Bipolar Transistor technology (SiGe-HBT), characterized by low series noise and thus best performance in terms of ENC (a few hundred electrons) for pulse-shaping times of less than 10ns.

## 2. Experimental setup of the test beam at the CERN SPS.

### 2.1 Sensor layout

The silicon sensors for this measurement were produced by ADVACAM[1]. They consist of 100μm thick planar n-on-p[2] with readout pads of 800μm × 800μm area and a pad spacing of 100μm. Each sensor had a total of 25 readout pads. Only one readout pad was read for this measurement, while the others were shorted and connected to the same voltage reference (Figure 1). The bias voltage could be applied both on the pads and the backplane through 100kΩ resistors.

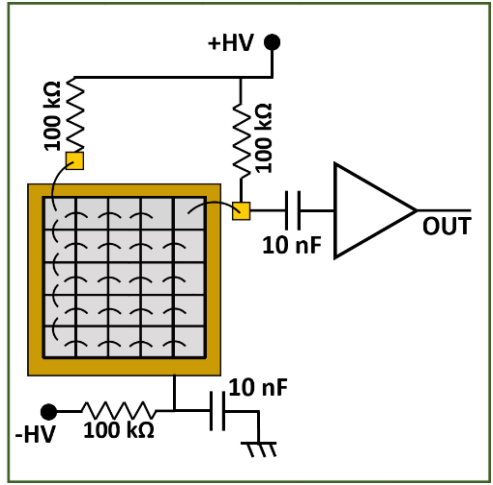

**Figure 1: Sensor polarization and readout scheme.**

---

[1] http://www.advacam.com/en/

[2] It should be noted that with a uniform field readout there is no difference between the n-on-p sensor signal and the signal of a p-on-n sensor, except for the pulse polarity.

## 2.2 Electronics performance

The preamplifier used for this test was produced using discrete component SiGe HBT transistors and based on a concept originally developed in silicon BJT to read diamond and RPC detectors [3]. Figure 2 shows the conceptual schematics of this amplifier.

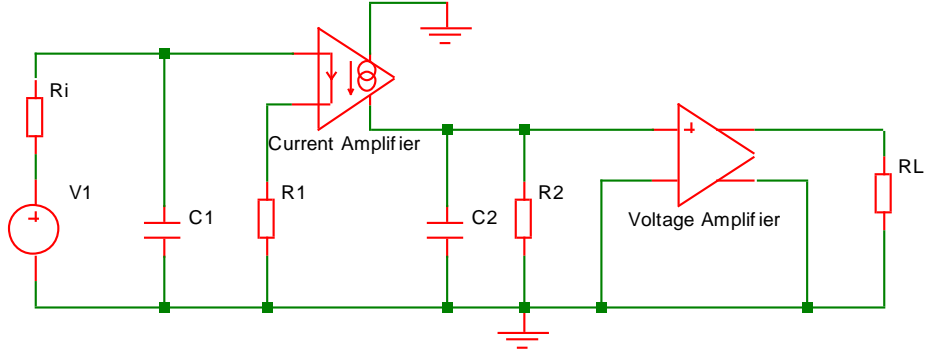

**Figure 2: Schematic representation of the preamplifier.**

The first amplification stage is a charge amplifier, designed to operate on an input capacitance of the order of 1pF. It operates as an ideal charge integrator (rise time equal to charge collection time) for pulses with duration down to 200ps, so that it does not exceed the charge collection time of a 100µm thick silicon sensor even at saturation of the carrier drift velocity. The ENC at 1pF input capacitance obtained from the simulation is 500 electrons RMS, allowing operation at a threshold of 0.5fC with pulses shorter than 1ns. The amplifier parameters are listed in Table 1.

**Table 1: Amplifier parameters.**

| Voltage Supply | 3-5 V |
|---|---|
| Sensitivity | 3.5-6.0 mV/fC |
| ENC (1pF input capacitance) | 500 e$^-$ RMS |
| Input Impedance | 100 - 200 Ohm |
| Bandwidth | 100 - 200 MHz |
| Power Consumption | 10 mW/ch |
| Rise Time δ(t) input | 200 - 400 ps |
| Radiation Hardness [3] | 50 Mrad, $10^{15}$ $n_{eq}$cm$^{-2}$ |

## 2.3 Test beam setup

Figure 3a shows a photograph of the silicon sensor assembly. The measurement of the time resolution of two of these silicon detectors was carried out at the H8 beam test facility at the CERN SPS, with a de-bunched pion beam of 180GeV energy.

The two sensors under test are as described in section 2.1; their thickness of 100μm allows the high energy pions to be considered Minimum Ionizing Particles (MIPs) for the distribution of energy deposited into the sensors.

The sensors were placed on two separate boards, together with the amplifier channel and later aligned by means of four precision pins and independently shielded. The boxes were supported by an aluminum plate suspended on three springs to allow micro-correction of the sensor alignment (Figure 3b).

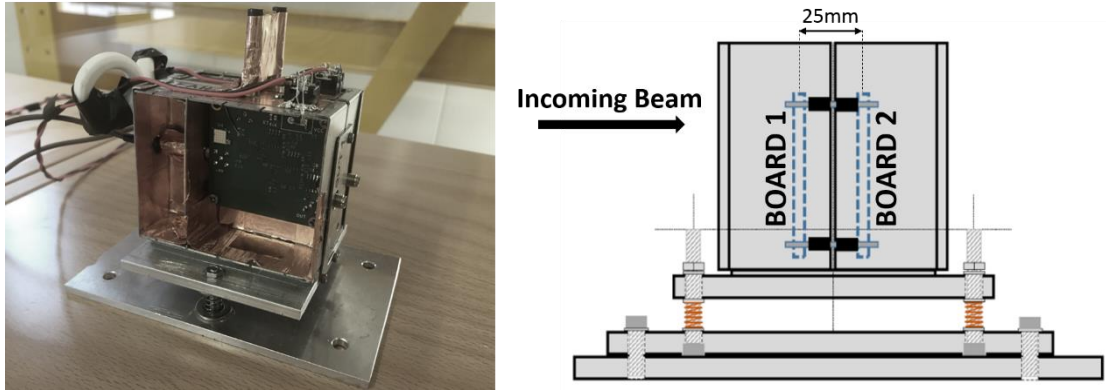

**Figure 3: a) Photograph of the assembly containing the two detectors. b) Layout of the support structure showing the position of the two sensor boards.**

The whole structure was fixed to a moving table controllable from outside the experimental area and aligned with the University of Geneva FEI4 pixel telescope [4], which provided an external trigger region of interest to the detectors under test (Figures 4 and 5).

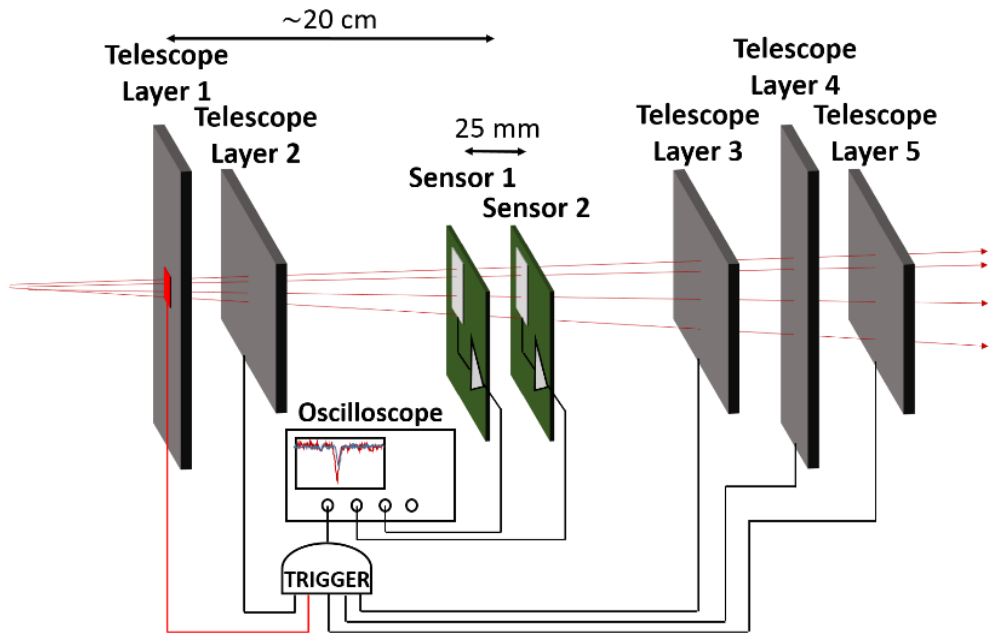

**Figure 4: Schematic representation of the test beam setup. The pion beam was coming from left side of the figure. The beam divergence is exaggerated. The trigger region of interest is set on the first telescope layer.**

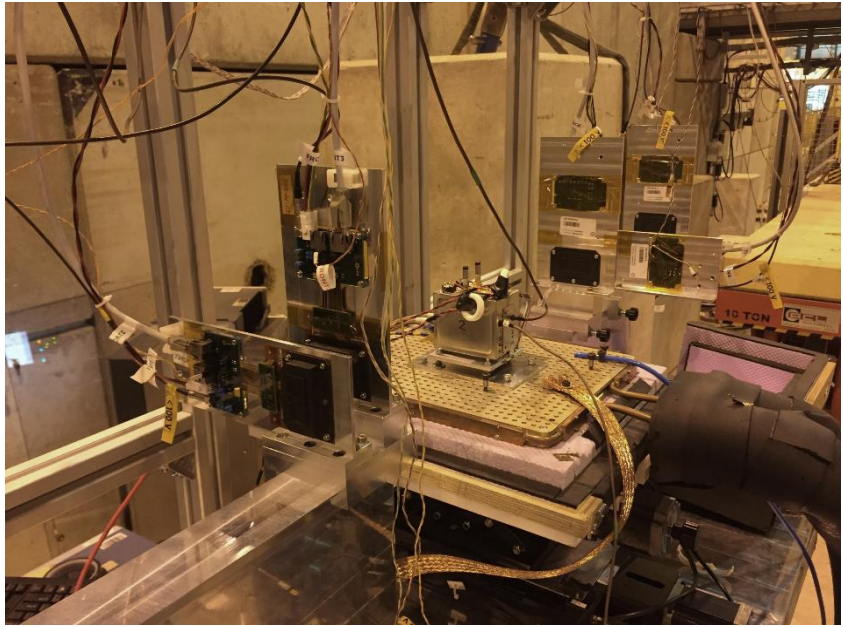

**Figure 5: Photograph of the test beam setup at the CERN SPS.**

The trigger was limited to an area of 500μm × 500μm on the first layer of the telescope, centered with the sensor pads that were readout.

The output signals of both amplifiers were sent to a Lecroy[3] WaveMaster 820zi oscilloscope by means of two 50Ω impedance coaxial cables and the waveforms were digitized in a 1μs window with a sampling rate of 40Gs/s. The oscilloscope bandwidth was limited to 1GHz for this measurement.

A negative bias voltage was supplied to the sensors from the backplanes with coaxial cables, while a DC value for all the pads was referenced to ground through 100kΩ resistors. Both amplifiers were operated at 4.5V.

Waveforms acquired by the oscilloscope were later analyzed by means of an analysis routine implemented in the ROOT software [5].

## 3. Analysis and results

### 3.1 Detector acceptance and efficiency

A total of 28536 events triggered by the external telescope were acquired and recorded at a bias voltage of 230V for both sensors under test.

Figure 6 shows a collection of pulses from sensor 2, from which the noise level and the rise time can be appreciated.

---

[3] http://teledynelecroy.com/

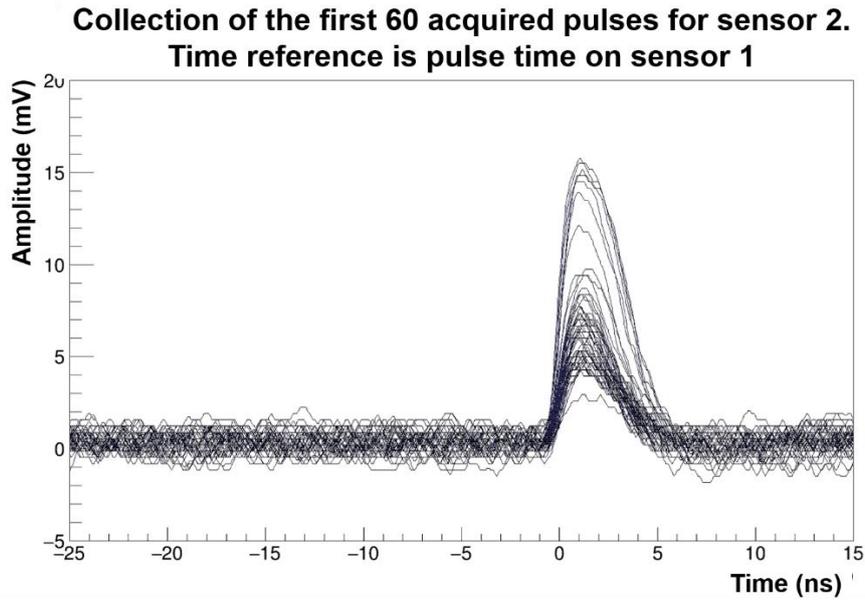

**Figure 6: Collection of pulses for sensor 2. The time reference is given by the pulse arrival time on sensor 1.**

The amplifier voltage noise (Figure 7) was measured by defining a background region of 600ns before the minimum pulse arrival time.

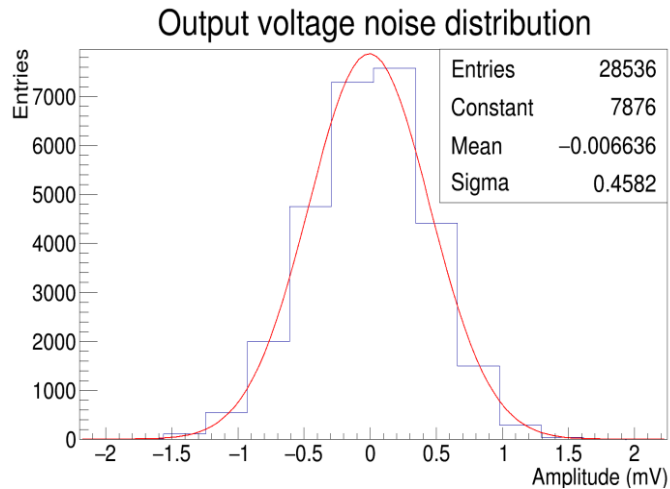

**Figure 7: Voltage noise at the amplifier output distribution.**

Despite the trigger being limited to a region of interest of 500μm × 500μm area centred in the sensor area, a slight misalignment of the sensors of about 100μm in both horizontal and vertical direction, together with the beam divergence, limited the detector geometrical acceptance to 95.6% for sensor 1 (upstream in the beam) and 93.4% for sensor 2.

In the calculation of the efficiency of each sensor, a signal of at least 3mV in the other sensor was requested in order to reduce the events that were outside the sensor geometrical acceptance due to the beam divergence.

The amplitude distributions of the output pulses for both sensors obtained with the selection explained above are shown in Figure 8.

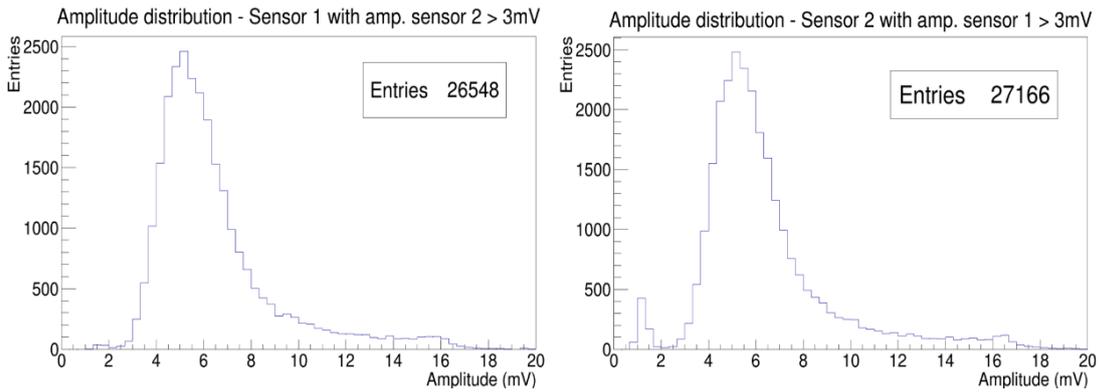

**Figure 8:** (Left) Pulse amplitude distribution for sensor 1, with event selection on sensor 2. (Right) Pulse amplitude distribution for sensor 2, with event selection on sensor 1. Sensor 2, downstream in the beam, shows a larger pedestal due to its lower geometrical acceptance from the beam divergence.

The product of the geometrical acceptance after the hit request on the other sensor and the detector efficiency can be evaluated by counting the fraction of events above a threshold of 2.3mV, equivalent to 5 RMS of the voltage noise. The result is a lower limit for efficiency of 99.7% for sensor 1 and 97.5% for sensor 2, being the latter limited by the beam divergence. With an expected most probable value of the deposited charge in a 100μm thick sensor of 6400 electrons for a MIP, the amplifier ENC is measured to be 540 electrons RMS, in good agreement with the calculated value of Section 2.2.

**3.2 Detector time resolution**

To verify that the amplifier behaves as a charge integrator, the distribution of pulse rise-times at 20%-80% of the amplitude is shown in figure 9 for sensor 2.

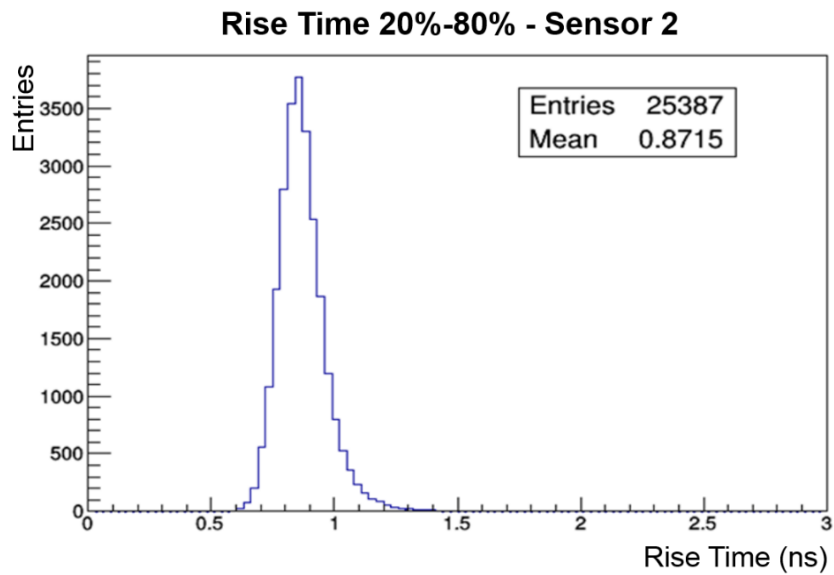

**Figure 9:** Pulse rise time (20%-80%) distribution for sensor 2, selecting signals with more than 3mV amplitude. The non-Gaussian tail at large values is due to the noise in small pulses when computing the 20% threshold value.

The plot in Figure 9 was obtained using a confidence window of 3ns for the rise time itself. The threshold at 20% is too low for the smallest pulses and is hardly distinguishable from the noise; this is the source of the non-Gaussian tail at large rise time values. The distribution is peaked at 850ps, as expected for an ideal charge integrator in a 100μm thick silicon sensor with uniform field operated at $2.3 V/\mu m$.[4]

To obtain the time resolution of the detectors, the arrival time of the pulses at a constant threshold of 2.3mV was measured. This time measurement is affected by the signal time walk, which produces a shift of the average value for different pulse amplitudes. To correct for the time walk, the average time difference between the detectors as a function of the pulse amplitude was measured for both sensors, from which polynomial correction functions were obtained (Figure 10).

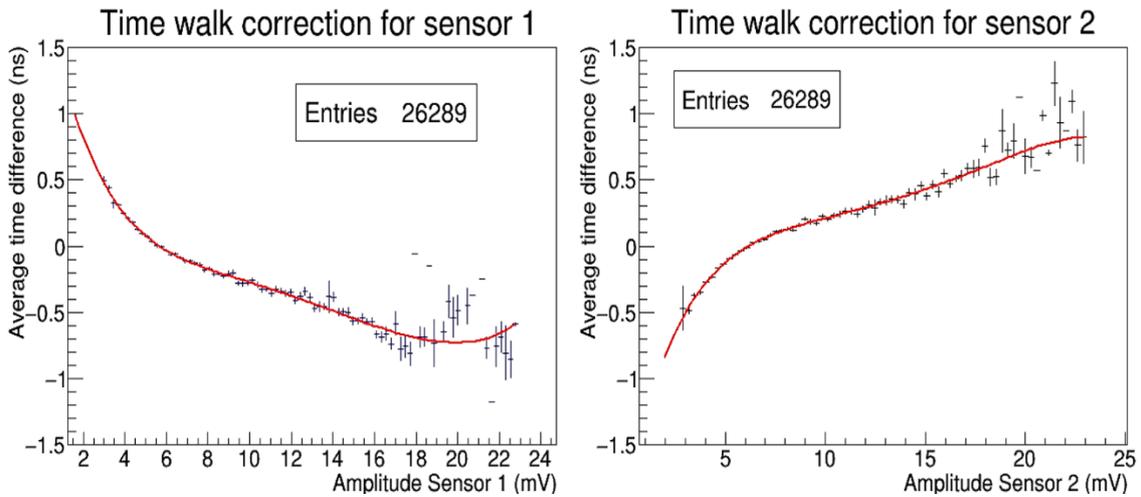

**Figure 10: The average pulse time difference as a function of the amplitude of sensor 1 (left) and sensor 2 (right). The pulse arrival time is evaluated at a fixed threshold of 2.3 mV in each sensor. The polynomial functions used for the correction are also shown.**

The distribution of the time difference between the two detectors, corrected for time walk and with the average set to zero, is shown in Figure 11. The convolution of the time resolution of the two detectors is the standard deviation of the Gaussian distribution, which is measured to be (150±1)ps.

---

[4] The electron and hole drift velocities in silicon at 2.3V/μm bias voltage are respectively 90μm/ns and 50μm/ns, so the collection time difference for 80% and 20% of the charge is expected to be between 0.7ns and 1.2ns.

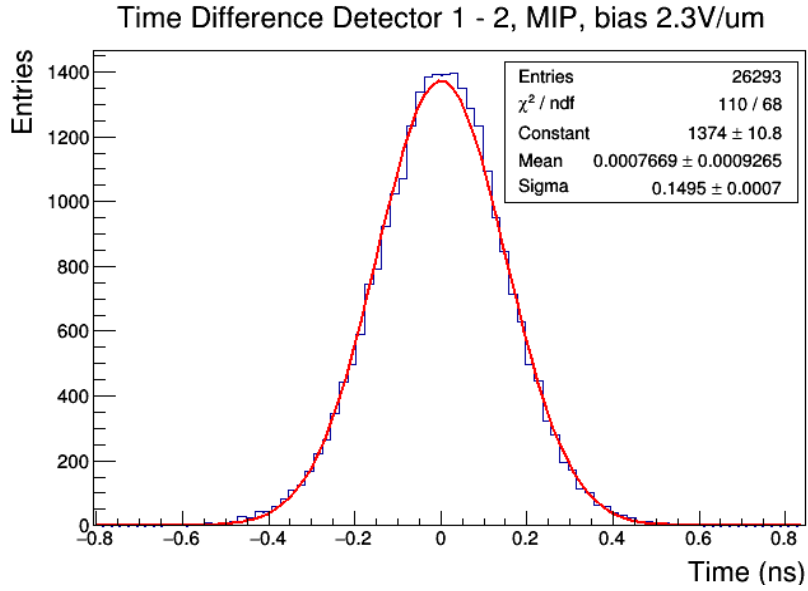

**Figure 11: Pulse time difference between the two detectors corrected for time walk effect and fitted with a Gaussian function. The expected mean value was not estimated and it was set to zero in the plot.**

Since the two detectors are in principle identical, as a first approximation the time resolution can be estimated as the standard deviation of the time difference divided by square root of two, thus obtaining:

$$\sigma_t = \frac{(150 \pm 1) ps}{\sqrt{2}} = (106 \pm 1) ps.$$

To compare the above result with the expectation for a pure voltage noise contribution to the time resolution, the distribution of the voltage noise divided by the output pulse slope is shown in Figure 12.

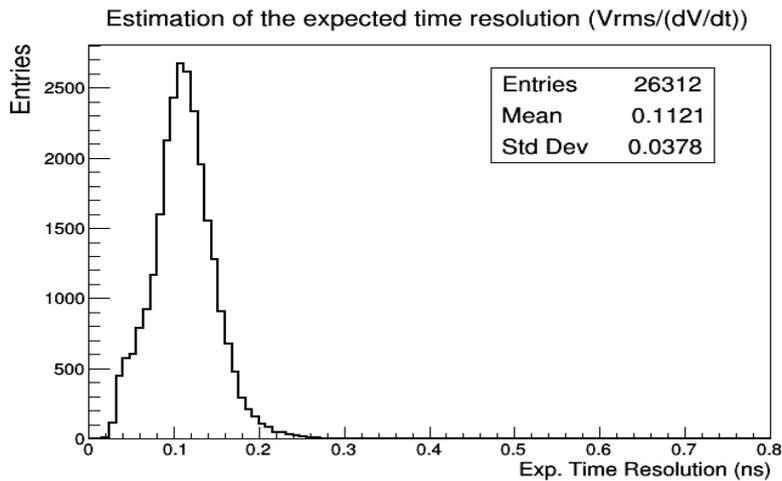

**Figure 12: Distribution of the voltage noise contribution to the theoretical time resolution for the acquired pulses. The mean value is a qualitative estimator of the expected detector time resolution.**

As already introduced in section 1.1, the best results in terms of time resolution are achieved by saturating the carrier drift velocity. To verify how the choice of the working point contributes to the timing performance of the sensor, special runs with less statistics were taken at lower values of the bias voltage. Figure 13 shows the measured time resolution as a function of the bias voltage.

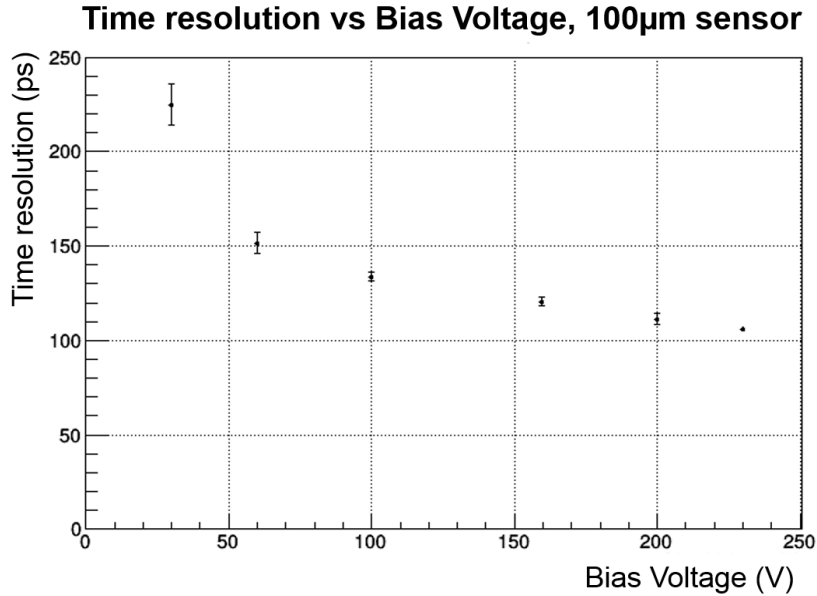

**Figure 13: Sensor time resolution as a function of the bias voltage. Full sensor depletion is not achieved for the first two points, and for this reason the time resolution performance could be overestimated due to the event selection.**

## 4. Conclusions

A time resolution of approximately 100ps for the detection of minimum ionizing particles has been measured with a 100μm thick p-on-n silicon sensor read out by SiGe HBT discrete component electronics. The sensor readout pad had a dimension of 800μm × 800μm and a capacitance of 1pF. The sensor layout and the electronics technology were carefully selected in order to achieve this result. No pulse fitting or multi-threshold techniques have been used. The time resolution obtained is compatible with our expectations.

## Acknowledgments

The authors want to thank Luigi Di Stante for his important contribution in designing and building the detector support structure, Gabriel Pelleriti for the assembling of the detector board and the staff of SPS North Area at CERN for providing excellent quality beam and infrastructure for the whole duration of the test. We are grateful to the colleagues who collaborated to the setup and operation of the test beam telescope and to the ATLAS-IBL collaboration for making available the FEI4 modules for the telescope. Finally, we wish to thank Allan Clark for many fruitful discussions and for reading this manuscript.